\documentstyle[prl,aps,epsfig]{revtex}
\title{
First Order
Nonequilibrium Phase Transition in a Spatially Extended System
}
\author{R. M\"uller, K. Lippert, A. K\"uhnel, and U. Behn}
\address{
Institut f\"ur Theoretische Physik, Universit\"at Leipzig, 
Augustusplatz 10, D-04109 Leipzig, Germany
}
\begin{document}
\maketitle

\begin{abstract}
We investigate a system of harmonically coupled identical nonlinear
constituents subject to noise
in different spatial arrangements.
For global coupling we find for infinitely many
constituents the coexistence of several ergodic components and
a bifurcation behaviour like in {\it first} order phase transitions.
These results are compared with simulations for finite systems both for
global coupling and for nearest neighbour coupling on
two- and three-dimensional 
cubic lattices. The mean-field type results for global
coupling provide a better understanding of the more complex behaviour
in the latter case.
\end{abstract}

The influence of noise on nonlinear systems is subject of intense
experimental and theoretical 
investigations \cite{moss}.
Zero-dimensional models considering
stochastic differential equations for a macroscopic order parameter 
homogeneous in space
exhibit noise induced transitions such as transitions
between unimodal and bimodal stationary distributions
\cite{schenzle,graham}, cf. also 
\cite{horsthe} and refs. therein. 
In real systems like solids and liquid crystals, however, 
interactions  exist between  spatially distributed constituents.
Spatially extended 
noisy systems
described by stochastic {\it partial} differential equations are
difficult to treat analytically, see, e.g. \cite{becker}.
Simulations --though expensive-- may provide some guide to a theoretical 
understanding of those systems 
\cite{broeck,broeck2,jing,pattern,park}.
\newline
Models with global coupling of nonlinear noisy constituents are
by far easier to investigate and allow even for analytical results
\cite{broeck,broeck2,dawson,hongler,shiino,jung}.
For example, Shiino \cite{shiino} was able to extend the concept
of phase transitions to nonequilibrium phenomena described by globally 
coupled nonlinear oscillators subject to additive noise. More recently,
Van den Broeck et al. \cite{broeck,broeck2} demonstrated the appearance
of a {\it second} order noise induced phase transition in a model
with globally coupled nonlinear constituents subject
to multiplicative and additive noise, which shows no
transitions in the absence of noise. In this paper we present
a model constructed in a  spirit similar to \cite{broeck} 
which exhibits a 
{\it first} order noise induced phase transition connected with
a hard onset of the coexisting ergodic components of the
system. Varying parameters of the system or of the noise
the order of the phase transition may change as we observed
previously for zero dimensional models \cite{behn}.\\
We investigate a system of harmonically coupled identical
nonlinear  constituents under the 
influence of noise acting simultaneously in additive and multiplicative 
way. We consider two cases distinguished by the spatial arrangement of
the $L$ constituents: The case of global coupling of all components
and the case of nearest neighbour coupling on a $d$-dimensional cubic 
lattice. In the case of global coupling, analytic results are obtained 
for    $L \rightarrow \infty$    and are compared
with simulations for $L=100$ and $L=1000$, respectively.
Furthermore, simulations were carried out for the case of nearest 
neighbour coupling in $d=2$ ($L=100 \times 100$) and $d=3$
 ($L=20 \times 20 \times 20$).
In these cases the results for global coupling can be considered
as mean field approximation.\\
The variables $x_i$ of the individual constituents at the lattice 
sites $i$ obey the following stochastic differential equations 
in the Stratonovich sense
\begin{equation}
\dot{x}_i  = f(x_i) +  g(x_i)\, \xi_i  - {D \over N} 
\sum_{j\in {\cal N}(i)} (x_i - x_j )\;.
\label{langevin}
\end{equation}
Here ${\cal N}(i)$ denotes the set of involved neighbours of 
site $i$ and $N = \# {\cal N}(i) $ is equal to $L-1$
in the case of global coupling and  to $2d$
in the case of nearest neighbour coupling. The parameter $D$ controls
 the strength of the spatial interactions.
The $\xi_i(t)$
represent zero mean spatially uncorrelated Gaussian white noise at 
point $i$ with autocorrelation function
\begin{equation}
\bigm < \xi_i(t)\, \xi_j(t') \bigm > \, =  
\sigma^2 \delta_{i j} \delta(t -t') \;,
\label{correl}
\end{equation}
where $\sigma^2$ is the noise strength. 
For nearest neighbour coupling and suitable chosen parameters, 
Eq. (\ref{langevin}) can be considered as discretized version of a 
stochastic partial differential equation with diffusive 
coupling.\newline
The stationary Fokker-Planck equation for the probability density
of $x_i$ reads \cite{broeck}
\begin{equation}
0  =  \frac{\partial}{\partial x_i} \bigg ( -f(x_i)+ 
\frac{D}{N}\sum_{j \in {\cal N}(i)}(x_i-\langle x_j | x_i \rangle) 
 +  \frac{\sigma^2}{2} g(x_i) \frac{\partial}
{\partial x_i} g(x_i) \bigg )
P_s (x_i) \;,
\label{fokker}
\end{equation}
where $\langle x_j | x_i \rangle = \int dx_j x_j P_s(x_j |x_i)$ is the 
steady state conditional average of 
$x_j$, $j\in {\cal N}(i)$ given $x_i$ at site $i$.\\
For the case of global coupling, fluctuations disappear in the average
$1/(L-1) \sum _{j \in {\cal N}(i)} \langle x_j | x_i \rangle$ if 
$L \rightarrow \infty$. 
Considering the class of solutions for which this expression is
independent of lattice site $i$ we can replace it by the steady 
state mean value    $\langle x \rangle $ which is 
self-consistently determined by
\begin{equation}
\left\langle x \right\rangle =  
\int_{-\infty}^\infty dx\,  x P_s \left( x,  \left\langle x 
\right\rangle \right)
\equiv F\left( \left\langle x \right\rangle \right)\;.
\label{selbst}
\end{equation}
Obviously, for finite lattices or coupling to a finite subset of 
neighbours this replacement represents a mean field approximation.\\
Following Shiino \cite{shiino} one obtains the same results by replacing
in (\ref{langevin}) the spatial average 
$ 1/N \sum _{j \in {\cal N}(i)} x_j $
by the statistical average $\langle x \rangle$.\\
In this paper we consider a simple model, for which a nontrivial
solution
of (\ref{selbst}) is not emerging from zero but appears with a jump 
to a nonzero value at a critical value
of the control parameter.  The model is
specified by
\begin{equation}
f(x) = ax + x^3 - x^5 \; ,\qquad    g(x) = 1 + x^2 \;.
\label{model}
\end{equation}
For the model with global coupling the stationary probability density is
         
\begin{equation}
P_s(x,\left\langle x \right\rangle)  \propto  
\big (1+x^2 \big )^{\textstyle \frac{3}{\sigma^2}-1} 
\exp \Bigg\{\sigma^{-2}
\bigg [-x^2+
\frac{D-a+2}{1+x^2} 
 + D \left\langle x \right\rangle \bigg (\frac{x}{1+x^2} + 
\arctan(x) \bigg )\!
\bigg ]\! \Bigg \}.
\label{meanf}
\end{equation}
Without spatial coupling ($D=0$) the model
shows the following bifurcation behaviour:\\
In the deterministic case $(\sigma^2 = 0)$ the stationary
solution undergoes a
subcritical bifurcation at $a_c =0$. The multiplicative
noise shifts the bifurcation threshold of the maximum of the
stationary probability density to $a_c = \sigma^2$.
For weak noise ($\sigma^2 < 1$) the bifurcation is subcritical whereas
for  $\sigma^2 \ge 1 $ it is supercritical.\\
In the following we mainly restrict ourselves for the sake of 
convenience to the case $\sigma^2 = 1$ where the noise intensity 
is just sufficient  to produce a change from the deterministic
subcritical bifurcation into a supercritical bifurcation.
Then the extreme values of the stationary probability density are
$x_{st}  = 0$ and, if $a>1$, $x_{st}  = \pm \sqrt[4]{a-1} $.\\
The global coupling $(D \neq 0)$
favours a coherent behaviour of the spatially distributed
components which is --in a sense-- an effect opposite to the noise and 
will 'restore' the subcritical bifurcation. Hence we expect to observe 
a first order nonequilibrium
phase transition.\\
The bifurcation behaviour of $\left \langle x \right \rangle$
is governed by the self-consistency condition (\ref{selbst}). 
Since for  our model
$F\left(\left \langle x \right \rangle\right)$ is an odd function of 
$\left \langle x \right \rangle$
we always have the solution $\left \langle x \right \rangle = 0$.
Moreover, pairs of new stable and unstable nonzero solutions 
may occur in certain parameter ranges.
Only stable solutions can be observed in simulations. 
We remark without proof that
$F'\left(\left \langle x \right \rangle \right) < 1 $ 
is sufficient for stability (cf. \cite{shiino}).
The existence of more than one stable solution leads to the 
existence of several corresponding
stationary probability densities $P_s(x, \langle x \rangle )$. 
Therefore, a phase transition breaking
the ergodicity of the system is expected in the case of global 
coupling.\\
The typical behaviour of 
$F\left(\left \langle x \right \rangle\right)$ 
for our model is sketched in Figure 1.
Whereas the model investigated by Van den Broeck et al.
\cite{broeck} exhibits a {\it second} order phase transition, 
we find a {\it first} order phase transition 
connected with a {\it hard} onset of the nontrivial stable 
solution of the self-consistency condition
for our model in a certain parameter range
where the model without spatial coupling exhibits only the 
trivial solution
$\langle x \rangle = 0 $.\\
The phase diagram given in Figure 2 confirms the intuitive picture 
drawn above: The spatial coupling favours coherent behaviour of the 
components acting thus opposite to the noise.
The critical value of $a$ is reduced with increasing coupling 
strength $D$ and above
a critical strength of $D$ the transition is of first order.\\
Figure 3 shows the different solutions of the self-consistency 
equation (\ref{selbst})
for the order parameter $\left \langle x \right \rangle$ leading to 
different ergodic components
of the system as a function of the parameter $a$,
the spatial coupling constant $D$,
and the noise strength $\sigma^2$, respectively. In all cases a 
hard onset of the nontrivial stable solutions can be observed.\\
It is instructive to compare the results obtained by solution of 
the self-consistency condition (\ref{selbst}) with simulations on 
{\it finite} globally coupled systems
of different size. We consider a parameter set 
($\sigma^2=1, D=25, a=-1.5$) where
three different stable solutions of (\ref{selbst}) exist.
For small systems ($L=100$) the ergodicity breaking is not perfect.
We still observe a few transitions between the
different 'ergodic components' due to large fluctuations of 
$\left \langle x(t) \right \rangle$
around its stationary values. The trajectory of the spatial average
$\langle x(t) \rangle _L = (1/L)
\sum _{i=1}^L x_i (t)$
is shown in Figure 4.
For larger systems ($L=1000$) the fluctuations become smaller and
the system remains very long inside
one of the ergodic components. In that case, there are practically no 
transitions.  The initial conditions determine which of the ergodic 
components is selected.\\
We also performed simulations of the stochastic differential equation
(\ref{langevin}) for nearest neighbour coupling
on a $3$-dimensional cubic lattice with $L=20 \times 20 \times 20$ 
sites and on a $2$-dimensional square lattice with $L=100 \times 100$
sites.
 We used the same set of parameters as in 
the  simulations for global coupling.
Qualitatively one gets a  behaviour very similar to the case
of global coupling.
In Figure 5 we compare the probability density $P_s(x)$ at one
lattice site given by (\ref{meanf}) for the case of global
coupling with the results of simulations for nearest neighbour 
coupling on the $3$-dimensional lattice.
Although for the finite system there is no {\it perfect} separation into 
different ergodic components, the 
trajectories remain very long  in the corresponding 
'basin of attraction'. The histograms obtained 
by sampling those trajectories follow very closely the probability 
densities of the ergodic components for global
coupling. The value of $x_s$ for nearest neighbour coupling is about
$10$ percent smaller than for global coupling.\\
Simulations on a
$2$-dimensional square lattice ($L=100 \times 100$) exhibit
a similar qualitative behaviour, the quantitative agreement
with the results of the globally coupled model is 
--as has to be expected-- less satisfying.
Figure 6 compares the order parameter as a function of $a$ and $\sigma^2$
for the 2-dimensional lattice with the results for global coupling. Although
the bistable region is smaller than in the case of global coupling it remains
no doubt that it exists. We remark that the fluctuations (indicated by the
error bars in Fig. 6) are larger for the states with $\langle x \rangle \ne
0$ than those for $\langle x \rangle = 0$ which is a clear indication of the
multiplicative nature of the driving process. As in the case of global
coupling, for the $L=100 \times 100 $ system the trajectory stays inside the
ergodic component selected by the initial condition for a very long time. No
jumps were observed in our simulations running typically over a time
$t=5000$. Jumps between the ergodic components induced by large
fluctuations are observed in simulations with smaller systems. With
increasing size they become less frequent, cf. Fig. 7.\\
In this paper we investigated a model which exhibits a {\it first} 
order nonequilibrium phase transition due to a hard onset of
the coexistence of several 
stable ergodic components of the system.
Other models which we investigate at present exhibit the same 
behaviour. In a different context, a system of coupled 
Duffing oscillators was used to caricature a liquid to crystal 
transition \cite{munakata}. Mean field theory yields a 
first order nonequilibrium transition, which is preserved including
additive noise.
We also found changes from second order
to first order transitions by tuning parameters of the noise or the
system. In any case, both the nonlinear terms and the interplay
between deterministic and stochastic effects determine the
order of the transition.
In previous work \cite{broeck,broeck2,jing,pattern,park} only 
second order  noise-induced nonequilibrium phase transitions have
been observed.\\
Our results may be of interest in the context of experimental
investigations in electrohydrodynamic convection in nematic liquid
crystals subject to thermal fluctuations (additive noise) and/or 
an external stochastic voltage (multiplicative noise).
There are experimental hints \cite{rehberg} that the first transition
from the homogeneous state to the structured state 
might be weakly hysteretic although the deterministic theory predicts 
a supercritical bifurcation.  \\
\acknowledgments
\noindent
Support by the Deutsche Forschungsgemeinschaft DFG under grant Be 1417/3 is
gratefully acknowledged.\\

\section*{Note added}

After completion of this work we got knowledge of a paper by S. Kim, S.H.
Park, and C.S. Ryn (Phys. Rev. Lett. {\bf 78}, 1616 (1997)) where a first
order noise induced transition on a different system of globally coupled
oscillators is described.\\

\newpage
\begin{figure}
  \epsffile{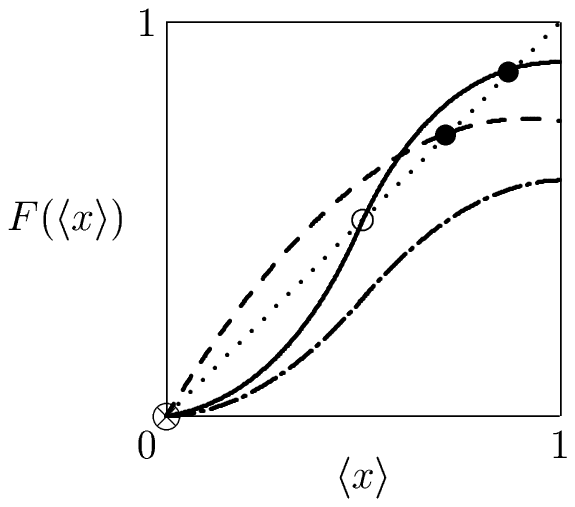}

FIG. 1.  Solution of the self-consistency equation (\ref{selbst}),
$F(\langle x \rangle ) = \langle x \rangle $, in three typical cases.
$\langle x \rangle = 0$
is always solution; in the case of the dashed-dotted line it is the 
only solution. In the
case considered by Van den Broeck et al. \cite{broeck} (dashed line)
we have two stable solutions
$\langle x \rangle = \pm x_s$ (full circle), $\langle x \rangle =0$ is 
unstable. In the case considered here
(solid line) we have besides the stable solution 
$\langle x \rangle = 0$ 
a pair of unstable solutions $\langle x \rangle = \pm x_u $ 
(empty circle) and a pair of
stable solutions $\langle x \rangle = \pm x_s $ (full circle). 
In contrast to the former case,
in the latter case the nontrivial solutions do not emerge continuously
from $\langle x \rangle =0$
but appear with nonzero value at the critical value of the control 
parameter.
This indicates a {\it first} order  nonequilibrium phase transition.
\end{figure}

\begin{figure}
\epsffile{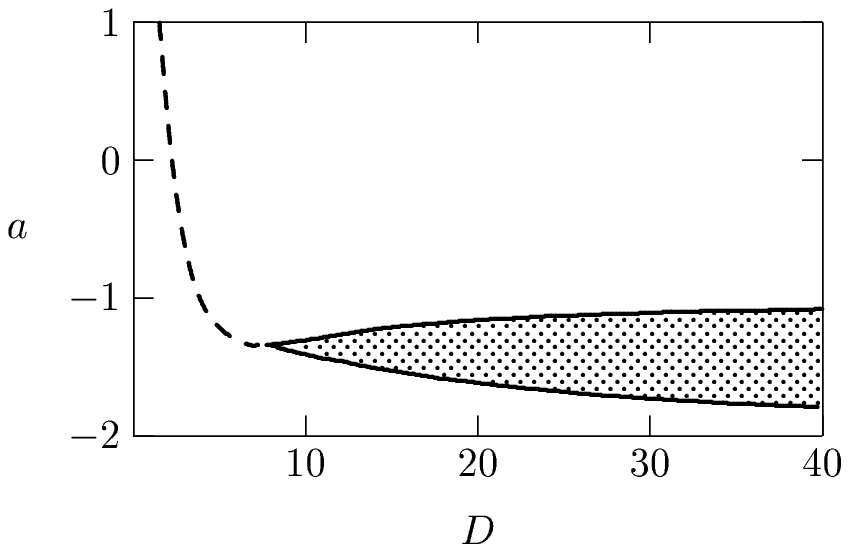}

FIG. 2.  Phase diagram in the case of global coupling for $\sigma^2 =1$.
For small $D$ we have a second order transition. The spatial coupling
favours a coherent behaviour of the constituents, acting thus opposite 
to  the noise. With increasing coupling strength $D$ the critical value 
of $a$ is reduced and
above a critical strength of $D$ the first order transition of the 
model without noise and spatial coupling is 'restored'. The solid 
(dashed) lines denote a first (second) order nonequilibrium phase 
transition. The number of ergodic components is three in the 
shadowed region, two in the region above and one in the region below. 
Hysteresis appears in the shadowed region.
\end{figure}

\begin{figure}
\epsffile{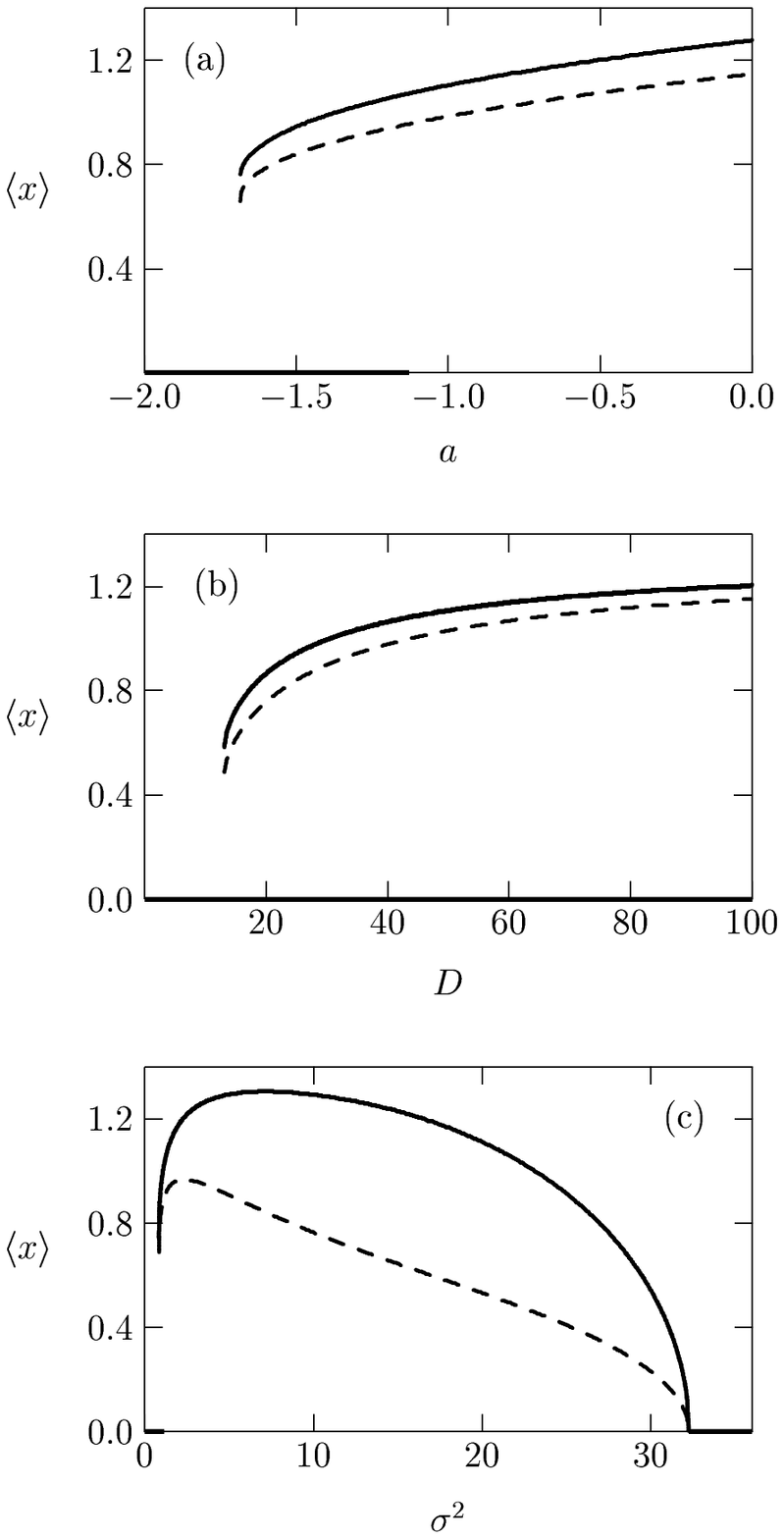}

FIG. 3.  Stable solutions of the self-consistency equation 
(\ref{selbst})
as function of the control parameters (solid lines) determine
the ergodic components. Unstable solutions (not shown here) cannot be 
observed in simulations. The maxima of the  stationary probability 
density (dashed line) of the corresponding ergodic components
(\ref{meanf}) exhibit qualitatively the same behaviour. In all 
the diagrams we observe at the critical value a {\it hard} onset of 
the nontrivial stable solution
corresponding to a {\it first} order transition. In (c) a reentrant 
behaviour is found similar to
that in \cite{broeck}. Parameters are $\sigma^2 =1$, $D=25$ in (a),
$\sigma^2 =1 $, $a=-1.5$ in (b), and $a=-1.5$, $D=25$ in (c).
\end{figure}

\begin{figure}
\epsffile{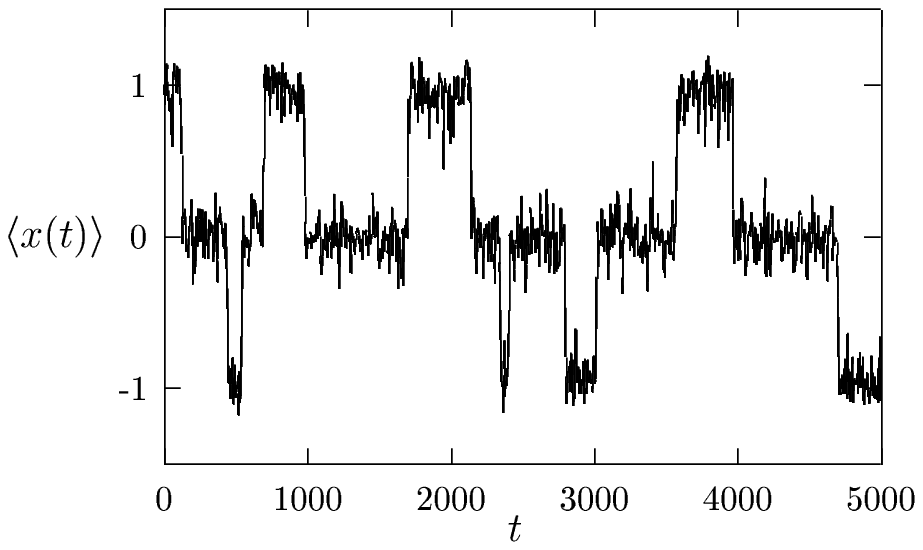}

FIG. 4.   Trajectory of the spatial average 
$\langle x(t) \rangle _L = (1/L)
\sum _{i=1} ^{L} x_i (t)$ for the case of global coupling 
($L = 100, \sigma^2=1, a=-1.48,
D=30 $). The trajectory fluctuates preferably around the mean values 
$x_s=0$ and $x_s= \pm 0.94$, sometimes large fluctuations lead to 
jumps between the 'ergodic components'.
\end{figure}

\begin{figure}
\epsffile{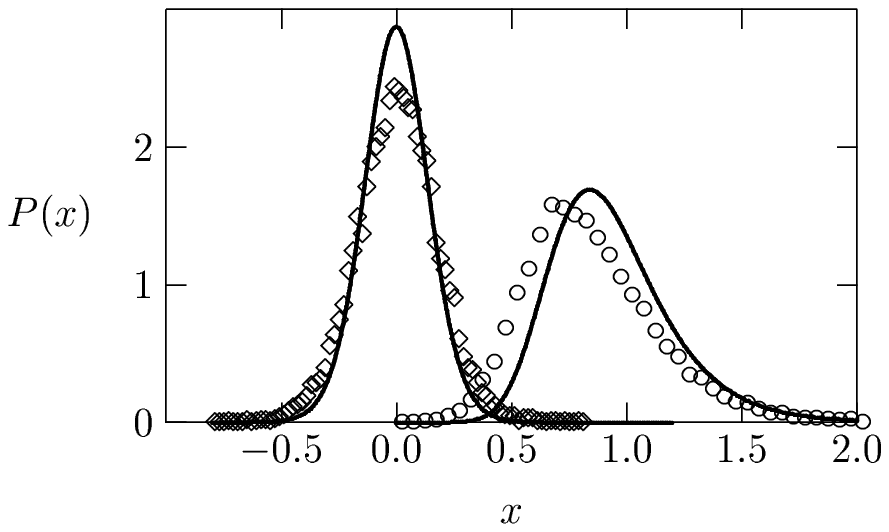}

FIG. 5.   Probability densities for $x_i(t)$ at arbitrary $i$
for the case of global coupling (solid lines) as given by
(\ref{meanf}). The ergodic components correspond to 
$\langle x \rangle = 0$, 
$x_s = 0.94$ ($\sigma^2 =1$, $D=25$, $a=-1.5$).
These results are compared with simulations for the 3-dimensional cubic lattice
($L = 20 \times 20 \times 20$)
with nearest neighbour coupling. The probability densities 
($\diamond$ and $\circ$) are
obtained by  sampling $20000$ equidistant points within a trajectory
of length $t=10000$ near $\langle x \rangle = 0$
and $x_s=0.85$ separately.
The plot indicates that the globally coupled system gives a good idea 
of the qualitative and quantitative behaviour of the system with 
nearest neighbour coupling.
\end{figure}

\vglue 0.8cm
\begin{figure}
\epsffile{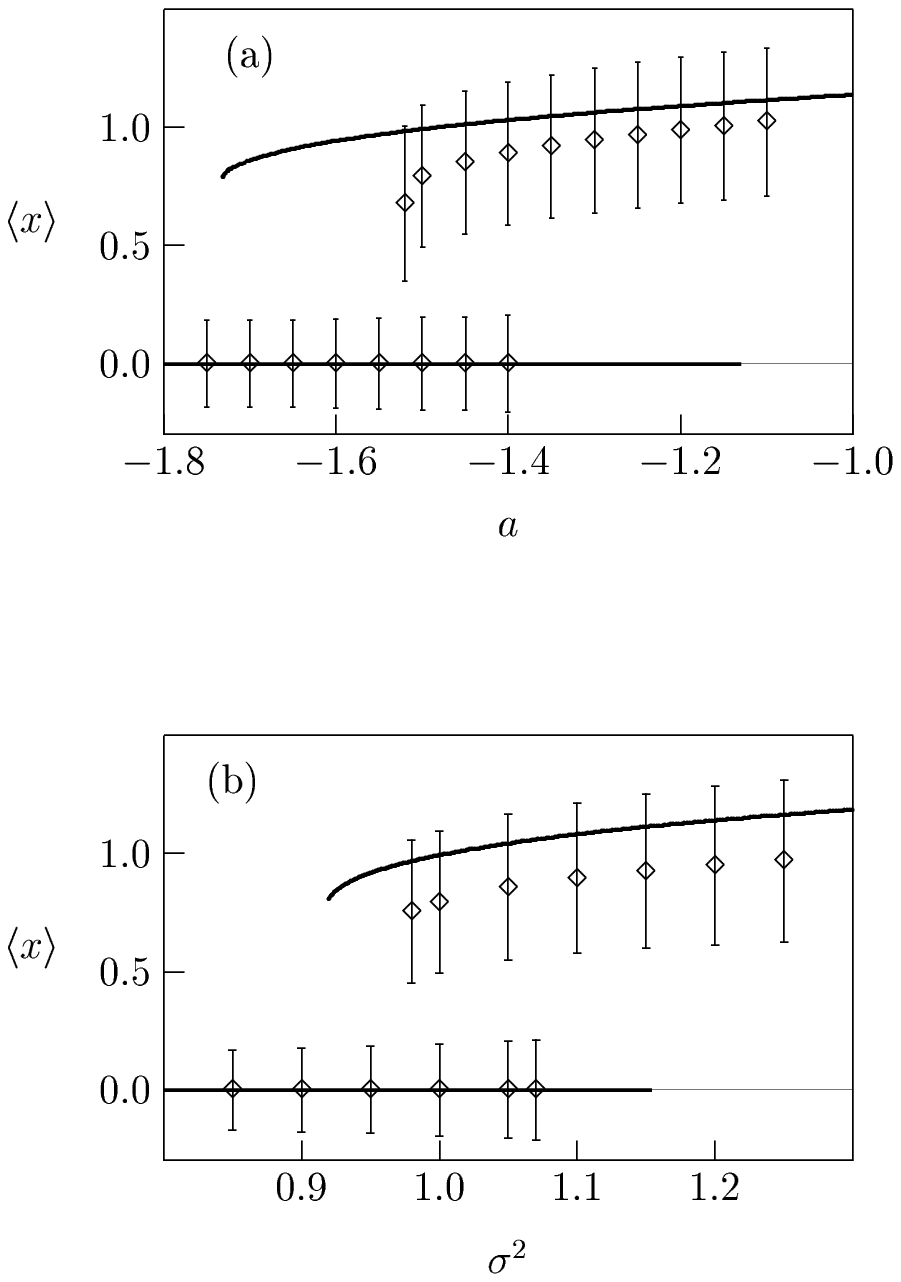}

FIG. 6. Comparison of the order parameter $\langle x \rangle$ obtained by
simulation for a 2-dimensional square lattice of size $L=100 \times 100$ with
the results for the globally coupled model (thick solid line) for $D=30$.
The diamonds
denote the average of $x_i(t)$ over all lattice sites and over a time span of
order  100 during which no jumps between the ergodic components occur. The
error bars indicate the time average over the standard deviation
$\left (1/L \sum_i (x_i(t) - \langle x(t)
\rangle _L)^2\right )^{1/2}$.
(a) and (b) show
the dependence on the control parameter $a$ ($\sigma^2 =1$) and the noise
strength $\sigma^2$ ($a=-1.5$), respectively. The coexistence of the
solutions with $\langle x \rangle \ne 0$ and $\langle x \rangle = 0$ over a
range of parameters as for the case of global coupling is obvious.
\end{figure}

\begin{figure}
\epsffile{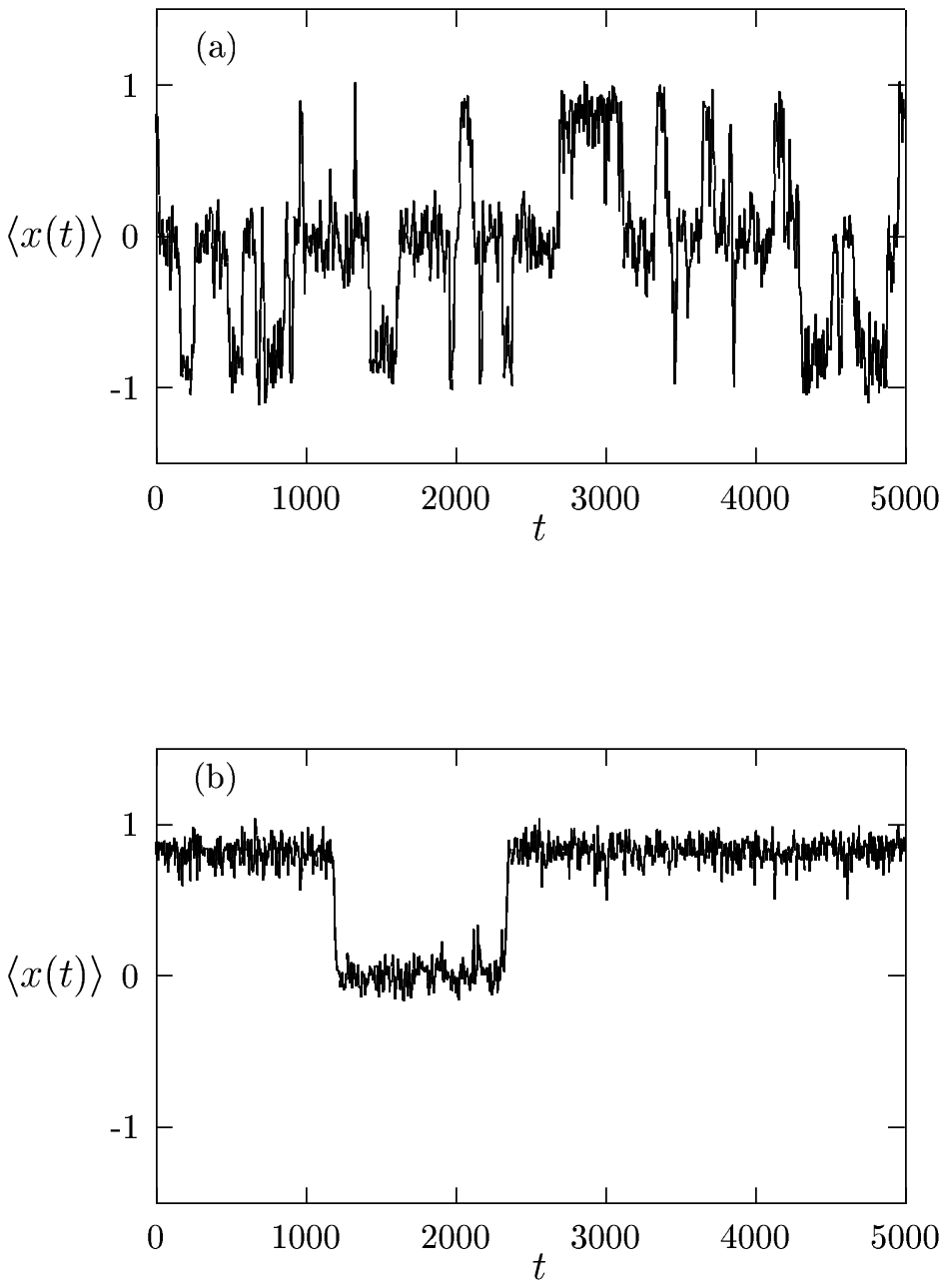}

FIG. 7. Trajectories of the spatial average $\langle x(t) \rangle _{L}$ for
the 2-dimensional square lattice of size (a)  $L=10 \times 10$  and (b)  $L = 18
\times 18$  for $D=30, \sigma^2 = 1, a=-1.48$. For the smaller system
frequent jumps between the 'ergodic components' are observed; with
increasing size of the system these events are rarefied. Already for a size
of $L=100 \times 100$ no jumps were observed in time spans of order 5000.
\end{figure}
\end{document}